# Suppressed Rupture of Thin Metal Films via van der Waals Epitaxy

Wenxiang Wang[1], Jiaxing Wang[2], Guotong Wang[1], Zhichao Yan[1], Chenxiao Jiang[3], Siqin Zhou[3], Chuanli Yu[1], Jianhao Chen[2], Kun Zheng[2], Thomas Salez[4], Xiaoding Wei[1], Zhaohe Dai[1,*]

[1]School of Mechanics and Engineering Science, State Key Laboratory for Turbulence and Complex Systems, Peking University, Beijing 100871, China.

[2]Beijing Key Laboratory of Microstructure and Property of Solids, Faculty of Materials and Manufacturing, Beijing University of Technology, Beijing, 100124, China.

[3]CAS Key Laboratory of Nanosystem and Hierarchical Fabrication, CAS Center for Excellence in Nanoscience, National Center for Nanoscience and Technology, Beijing 100190, China.

[4]Univ. Bordeaux, CNRS, LOMA, UMR 5798, F-33400 Talence, France.

*Corresponding author. Email: daizh@pku.edu.cn

Ultrathin metal films exhibit liquid-like instabilities, rupturing via surface diffusion far below their melting points. This behavior constrains thermal budgets for advanced integrated circuits and emerging 2D-crystal devices. Here, we demonstrate that these instabilities can be fundamentally suppressed using graphene as a van der Waals (vdW) template. While conventional 20-nm-thick gold films break up into islands below 300 °C, templated films not only remain stable but also become structurally refined after annealing above 600 °C. This exceptional stability stems from a vdW-mediated crystallographic texture that reorganizes grain boundaries into a mechanically robust network. This mechanism significantly widens the processing window for nanoscale interconnects and enables high-temperature integration of metals with 2D-crystal technologies.

Liquid films on a non-wetting surface spontaneously rupture and break up into droplets through dewetting (1). Solid films, both crystalline and amorphous, exhibit similar instabilities at sufficiently small scales or elevated temperatures (2-4). Nanometric metal films, such as Au, are particularly vulnerable; their as-deposited microstructures are far from equilibrium, making them prone to rapid breakup (5). The rupture temperature scales inversely with film thickness and can be as low as one-tenth of the bulk melting point ($T_m$) (5). While this instability has long challenged the fabrication of traditional integrated circuits and microsystems (6), it has become even more problematic in emerging 2D-material technologies, where ultrathin metal layers (<50 nm) serve as contacts, gates, and interconnects, and where annealing is essential to form low-resistance metal–2D junctions (7-22). Consequently, current fabrication protocols largely restrict processing to below 300 °C (Fig. S1), well below the ~400 °C thermal budget required for standard semiconductor integration (12, 13, 23, 24). Here, we report the exceptional thermal stability of ultrathin metal films achieved by van der Waals (vdW) epitaxy. For example, a 40-nm-thick Au film not only survives but exhibits enhanced structural integrity after annealing at 850 °C (0.8 $T_m$).

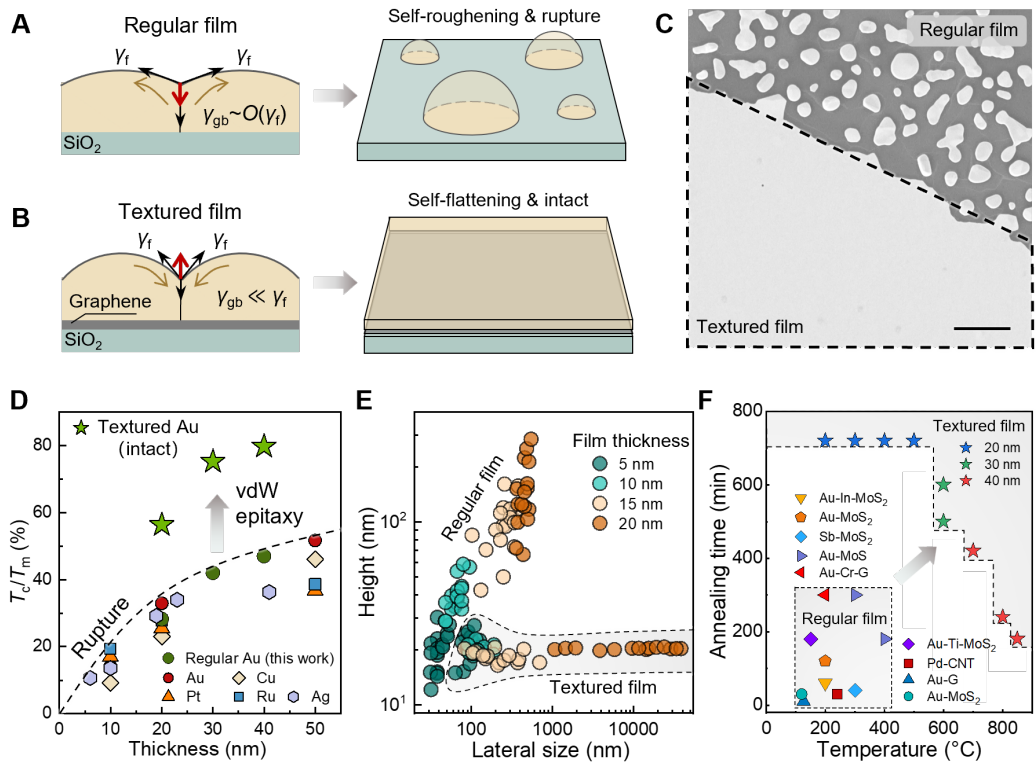

**Fig. 1. Evolution of solid films at elevated temperatures. (A)** Schematic of a conventional polycrystalline metal film (regular film), in which surface diffusion drives self-roughening and rupture. **(B)** Schematic of a graphene-templated film (textured film), where surface flow is directed toward self-flattening rather than breakup. **(C)** SEM image of a 20-nm-thick Au film annealed at 600 °C for 2 hours; the dashed box highlights

the textured region. Scale bar, 2 μm. **(D)** Critical rupture temperatures for conventional metal films (lower markers; data reproduced from Refs. (25) and (26)) and temperatures at which textured Au films remain intact (upper markers). All temperatures are normalized by the melting point. **(E)** Height and lateral dimensions of Au films after annealing at 600 °C for 2 hours. Note that textured films only break up at small thicknesses and begin to flatten when their thickness exceeds 20 nm, such that the lateral size is controlled by the size of the underlying graphene sheet (> 1 μm). **(F)** The processing window applied to typical 2D material devices in the literature, including Au-In with $MoS_2$ (7), Au with $MoS_2$ (8), Sb with $MoS_2$ (12), atomic layer bonding (Au-$MoS_2$) (13), Au with 2D materials with Cr or Ti adhesion layer (16, 27), Pd with carbon nanotubes (17), Au with graphene (18), and transferred Au with $MoS_2$ (22). Star markers represent specific conditions where textured Au films are found remain intact.

**From dewetting to capillary leveling**

We fabricated Au films with thicknesses ranging from 5 to 40 nm using standard electron-beam evaporation (See Materials and Methods). Films deposited directly onto $SiO_2$ substrates are referred to as "regular" films, while those deposited onto $SiO_2$ covered by a mechanically exfoliated graphene monolayer are termed "textured" films. Atomic force microscopy (AFM) reveals that the textured films exhibit slightly higher roughness than the regular films (Fig. S2C), as illustrated in Fig. 1, A and B. This behavior aligns with prior reports on Au on 2D surfaces (28), possibly due to the slightly modified effective surface energy by the atomic layer coating (29). Because thin-film deposition yields far-from-equilibrium microstructures, both types of films evolve upon heating (5). However, their behaviors differ dramatically, as seen in the scanning electron microscope (SEM) image in Fig. 1C. For example, after annealing at 600 °C for 2 hours, 20-nm-thick regular films rupture into isolated islands, whereas textured films remain continuous (and become even flatter than in the as-deposited state).

The rupturing behavior of regular films has been widely reported across many metallic systems (5). For ultrathin films (e.g., Au, Ag, Cu) with thicknesses of 5–50 nm, the critical failure temperature typically lies between ~10% and ~50% of $T_m$ (Fig. 1D), and our measurements fall within this range (25, 26). Mullins showed that a planar metallic surface is stable with respect to all perturbations (30). However, as Srolovitz and Safran clarified (31), polycrystalline metallic films almost inevitably develop grooves where grain boundaries intersect the free surface (Fig. 1A). In this configuration, the disjoining pressure is not important, while the grain boundary energy $\gamma_{gb}$ can be comparable to surface energy $\gamma_f$, producing a net driving force to deepen the groove (14). Surface diffusion then transports material away from the groove, nucleating holes and eventually leading to rupture at relatively low temperatures (32).

By contrast, the flattening of textured films is unexpected. Particularly, the textured films are slightly rougher (Fig. S2C), and the atomically smooth graphene surface provides lower friction (33, 34); These conditions would normally promote surface diffusion and accelerate rupture (35). Indeed, previous studies have shown that ultrathin Au films (< 5 nm) form more mature island patterns on graphene than on $SiO_2$ (36, 37). Our measurements show that textured films readily break up when their thickness is below ~10 nm (Fig. S2B), although the resulting islands are shallower and more puddle-like than those formed by regular films (Fig. 1E). Crucially, once the textured films exceed ~20 nm, they no longer rupture at 600 °C but instead evolve toward an increasingly leveled morphology (Fig. 1E).

The anomalous leveling behavior allows 40-nm-thick textured Au films to remain fully intact at 850 °C for 3 hours, approaching 80% of $T_m$ (Fig. 1D). This stability is remarkable given that even 50-nm-thick single-crystal Si films begin to rupture at 800 °C after only 2 minutes (38). Moreover, the 20-nm-thick textured films are intact at 600 °C and, even at 800 °C, exhibit only partial edge retraction while remaining continuous across their interior (Fig. 2B). Such excellent thermal stability is not limited to monolayer graphene templates: similar behavior is observed when using other 2D material templates, including exfoliated multilayer graphene, $MoS_2$, and large-area chemical vapor deposition (CVD) grown graphene (figs. S3-S4), indicating a universal fundamental mechanism.

To avoid the thermal failure of thin-metal contacts, gates, and interconnects (39, 40), most 2D-device fabrication protocols operate within a narrow thermal budget—typically below 300 °C for ~300 min (Fig. 1F) (7-22). A notable exception is a recently reported "atomic layer bonding" approach that removes the top chalcogen layer of $MoS_2$, enabling annealing up to ~400 °C (13). By contrast, our results show that textured metallic films can extend the accessible process window to temperatures approaching 800 °C and annealing times of ~800 min (Fig. 1F), while remaining continuous and even exhibiting improved microstructure and electrical conductivity (to be shown in Fig. 4).

**Suppressed edge retraction**

The results in Fig. 1C indicate that textured films can exhibit thermal stability not only in their interior regions but also at their edges. This behavior is also unusual, as film edges typically retract during annealing, and the time dependence of the retraction distance has been used to extract the diffusivity of solid films (41-43). To clarify this edge behavior, we further fabricated 20-nm-thick stepped Au films (Fig. 2, D and E) and monitored their edge evolution during annealing at 300 °C (44). We find that textured films exhibit only slight edge retraction at early times (Fig. 2D and Fig. S5), whereas regular films show both retraction and rupture near the edge (Fig. 2C). Remarkably, edge retraction in textured films ceases entirely after approximately 1 hour of annealing, as further evidenced in Fig. 2F, which plots the retraction length as a function of time.

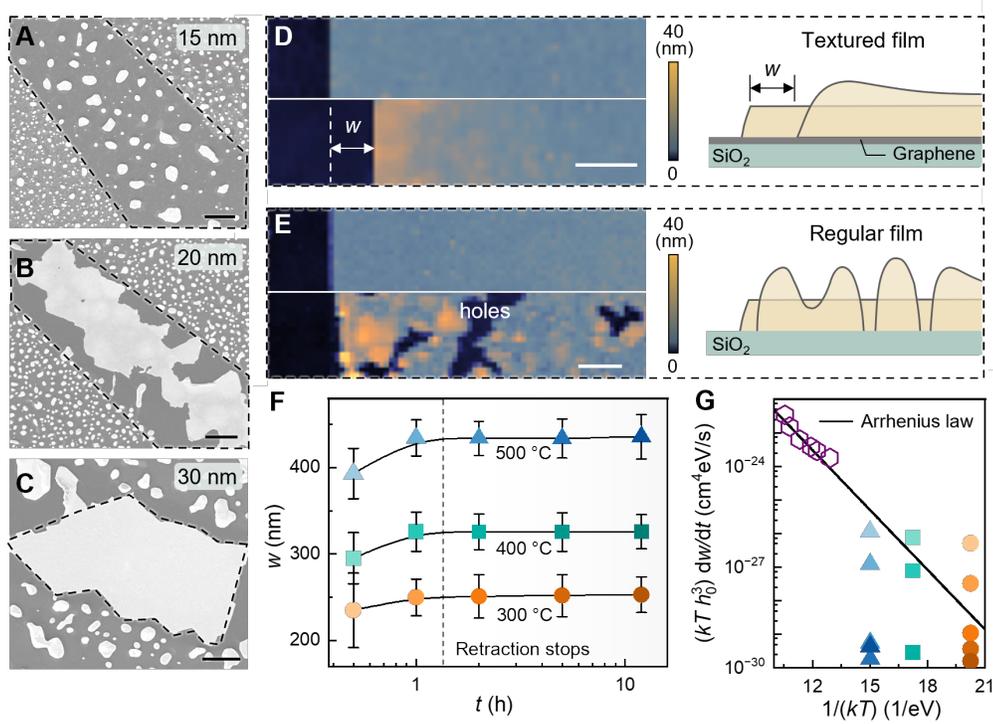

**Fig. 2. Evolution of stepped Au films.** (**A-C**) SEM images of Au films with different thicknesses after annealing at 800 °C for 2h; dashed boxes denote textured regions. Scalebars, 2 μm (A), 5 μm (B), 3 μm (C). (**D** and **E**) Schematic illustrations and AFM images of a stepped textured (D) and regular (E) film before and after annealing at 300 °C for 30 min. (**F**) Edge retraction length of the stepped textured film as a function of time, measured at different annealing temperatures. (**G**) Relation between the retraction speed and annealing temperature. The solid line and open markers represent data for regular Au films adapted from Ref. (43).

The retraction length w of textured films evolves with time t in a manner that differs qualitatively from prior measurements on thicker, regular films (with unperturbed thickness $h_0 \geq$ 40 nm). In particular, the early-stage dynamics (t < 1 h) exhibit a temperature dependence that departs from the Arrhenius law $dw/dt \propto \exp(-E_a/kT)/kT$ reported by Jiran and Thompson (Fig. 2G), where k is Boltzmann's constant, T is the absolute temperature, and $E_a$ is the activation energy (43). Furthermore, because retraction arrests rapidly, the previously observed steady-state scaling, $w \propto t^{2/5}$, was not evidenced in textured films (41, 42). We note that this arrest is not dictated by the graphene interface alone; Indeed, a regular film that is peeled off and subsequently transferred onto a graphene-coated substrate still ruptures upon annealing (Fig. S6). We therefore attribute the exceptional thermal behavior of textured films to their distinctive as-deposited microstructure, the details of which are clarified below.

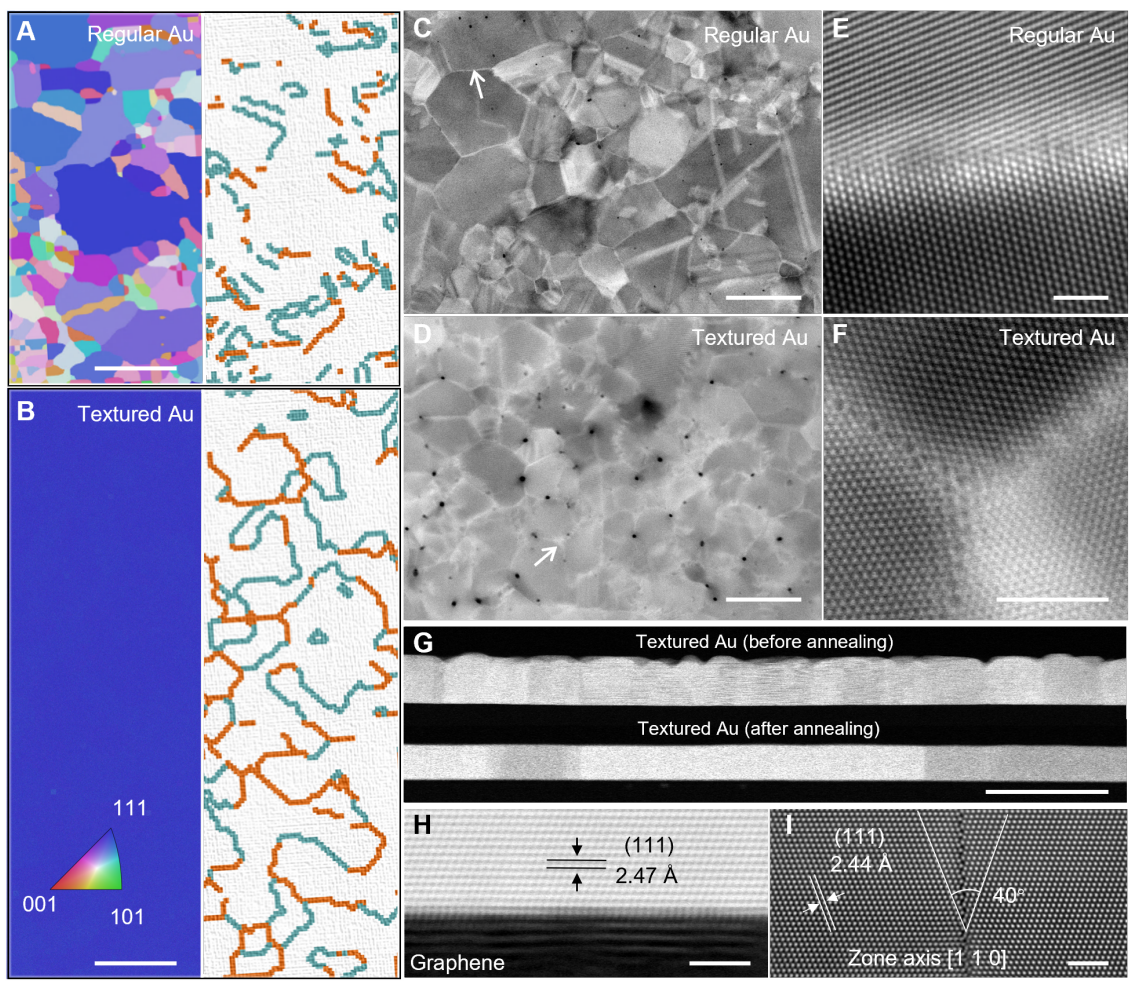

**Fig. 3. Microstructural characterization of Au films.** (**A** and **B**) EBSD images showing grain orientation and grain-boundary features of regular (A) and textured (B) Au films; Σ3 boundaries are highlighted in green and other coincident site lattice boundaries in orange. (**C** and **D**) Plane-view HAADF-STEM images of regular (E) and textured (F) Au films. (**E** and **F**) High-magnification plan-view images of representative grain boundaries in regular (G) and textured (H) films. (**G**) Cross-sectional HAADF-STEM images of a 30 nm thick textured Au film before and after annealing at 600 °C. (**H**) Cross-sectional images of the textured Au/graphene interface after annealing. (**I**) Atomic-resolution cross-sectional image of a Σ3(112) twin boundary in the textured Au film after annealing. Scale bars, 100 nm (A–C, G), 50 nm (D), 1 nm (E), 2 nm (F, H, I).

**A low-energy grain boundary network**

The crystal textures of the Au films, as determined by transmission electron backscatter diffraction (EBSD), are shown in Fig. 3, A and B. While regular Au films exhibit random grain orientations, textured films display a strong out-of-plane (111) texture induced by the graphene template during deposition (Fig. S7). Additionally, regular films possess a broad grain-size distribution (10–190 nm) with a preponderance of grains in the 10–30 nm range (Fig. S8A). In contrast, textured films exhibit a significantly narrower and more uniform grain-size distribution (35–60 nm) (Fig. S8A). Molecular dynamics (MD) simulations (45-47) further reveal that Au atoms have much longer surface diffusion lengths on graphene than on $SiO_2$, promoting (111) growth and reducing grain density (Fig. S9). As a result, the van der Waals interaction with graphene drives the formation of an epitaxial architecture with uniform texture that is inherently resilient to thermal perturbation.

The structural uniformity of textured films gives rise to a unique network of mechanically robust grain boundaries (Fig. 3, A and B). Specifically, about 40% of these boundaries are low-angle (low-energy), and the remaining high-angle grain boundaries consist almost entirely of coincidence-site lattice (CSL) boundaries characterized by superior atomic alignment (Fig. S8B) (48-50). This suggests that nearly all grain boundaries in textured films possess low interfacial energies, thereby enhancing mechanical stability. In contrast, only 20% of the grain boundaries in regular films are low-angle (Fig. S8B). While some CSL boundaries are present in regular films, they appear only as short, dispersed segments that do not form a continuous network (Fig. 2A). This microstructural heterogeneity results in a high fraction (approximately 50%) of high-energy, large-angle grain boundaries (Fig. S8B), which function as rapid diffusion pathways under thermal perturbation. This stark contrast explains the opposing net line forces at the grain-boundary grooves; such forces drive rupture in regular films (Fig. 1A) but promote leveling and healing in textured films (Fig. 1B).

We further elucidate the difference in grain size and orientation between regular and textured films via plane-view, aberration-corrected high-angle annular dark-field scanning transmission electron microscopy (HAADF–STEM) in Fig. 2, C–F. Consistent with the EBSD analysis, atomic-resolution imaging shows that the textured film retains out-of-plane (111) orientations at triple junctions (Fig. 1F), while the grains in the regular film are misoriented at bi-junctions (Fig. 1E). These misoriented junctions form higher-energy boundaries that serve as preferential sites for void nucleation and subsequent rupture (31).

Interestingly, many tiny voids (radii <5 nm) are already present in the as-deposited textured films (Fig. 3D), likely resulting from the incomplete coalescence during late-stage Volmer–Weber growth (28, 51). Upon annealing, however, these voids do not expand; instead, they contract and heal through capillarity-driven mass transport (Fig. S10). This process is energetically favorable as it reduces the high-curvature Au free surface and minimizes the exposed Au/graphene interfacial area (32). Simultaneously, the low-energy grain-boundary network acts as a "reinforcing skeleton", effectively suppressing the onset of rupture. Note that electron energy-loss spectroscopy shows that such metal-graphene interaction is predominantly physical in nature (Fig. S11), providing a route for producing thermally robust, ultrathin, ultraflat, textured, and transferable metal films (52).

The surface of textured films becomes markedly more planar during annealing (Fig. 3G). The strong out-of-plane (111) texture is preserved (Fig. S12), and individual grains exhibit improved uniformity with a near-absence of discernible defects (Fig. 3G). High-resolution imaging reveals atomically ordered structures near the graphene interface (Fig. 3H) and sharply defined vertical incoherent twin boundaries (Fig. 2I). Notably, the epitaxial interaction between the ordered Au lattice and graphene can produce moiré fringes (Fig. S10), underscoring the high degree of structural refinement in the annealed textured films. In contrast, the high grain-boundary energy inherent in regular films leads to rupture during annealing, forming discrete nanoparticles in which numerous defects, such as dislocations and coherent twins, arise to dissipate the strain generated during morphological evolution (Fig. S13).

**Thermally robust devices**

The exceptional thermal stability of textured films provides a robust safety margin for the high-temperature annealing required for semiconductor devices. To demonstrate this, we examined metal gates, which can be as narrow as ~100 nm and are therefore particularly vulnerable to thermally driven instabilities (20, 53). We fabricated regular, textured, and Ti-adhered (3-nm) Au ribbons (40-nm thick) with widths ranging from 100 nm to 6 μm using electron-beam lithography. Electrical transport was characterized via four-terminal measurements (Fig. 4A).

As illustrated in Fig. 4B, conventional metal ribbons succumb to Rayleigh-Plateau instability upon annealing (54). In stark contrast, 100-nm-wide textured ribbons remain structurally intact after annealing at 700 °C (Fig. S14), whereas regular ribbons rupture near 400 °C, even with a Ti adhesion layer (Fig. 4C). Electrical measurements across all ribbon widths corroborate this morphological stability (Fig. 4D and Fig. S15). For identical geometries, textured ribbons show the lowest initial resistivity and, notably, a further reduction in resistance after annealing, approaching the bulk value of Au. This improvement is consistent with microstructural refinement—rather than degradation—of the textured films.

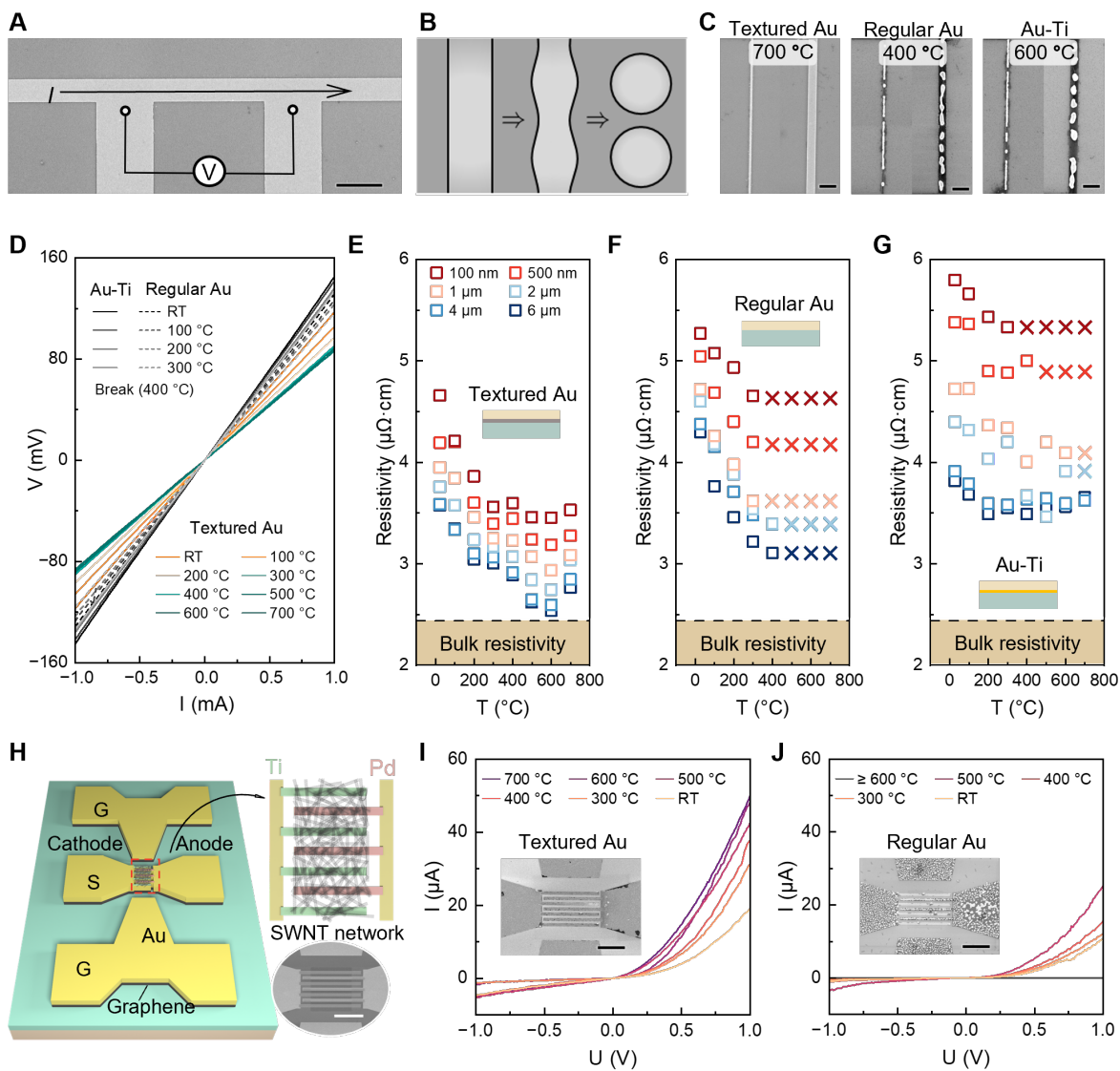

**Fig. 4. Electrical performance of textured Au ribbons.** (**A**) Four-terminal method for measuring the resistance of metal ribbons. Scale bar, 4 μm. (**B**) Schematic illustrating the Rayleigh-Plateau instability of thin ribbons. (**C**) Morphologies of textured, regular, and Au-Ti films with linewidths of 100 nm and 500 nm after annealing at 700 °C, 400 °C, and 600 °C, respectively. Scale bar, 1 μm. (**D**) Current-voltage curves of 100 nm wide ribbons after annealing at different temperatures. (**E-G**) Resistivity of textured (E), regular (F), and Au-Ti (G) ribbons of different widths after annealing at various temperatures. The symbol "×" indicates damaged ribbons due to Rayleigh-Plateau instability. (**H**) Schematic diagram of a radio frequency diode device using textured Au films for electrodes and connection lines. Ti metal serves as the cathode contact, textured palladium as the anode contact, and semiconductor single-walled carbon nanotubes as the channel layer. The lower-right corner shows the SEM morphology of the device. (**I** and **J**) Current-voltage

curves of carbon nanotube diodes based on textured (I) and regular (J) films at different temperatures. Scale bar, 20 μm.

We further demonstrate device-level robustness using large-area CVD graphene as a texturing template for Au electrodes and interconnects. Conventional $MoS_2$ transistors typically lose functionality after annealing at ~400 °C due to contact failure (even when Bi or Sb adhesion layers are employed) (12), transistors integrated with textured Au remain operational up to 600 °C (Fig. S16). We also integrate textured electrodes with thermally resilient single-walled carbon nanotubes (55, 56) to form radio-frequency (RF) diodes, which could experience substantial localized Joule heating during operation in the sub-6-GHz 5G operation (Fig. 4H). In high-temperature endurance tests, these devices not only operate normally after annealing at 700 °C for 2 h but also exhibit a further decrease in resistance (Fig. 4I). In contrast, devices using regular electrodes degraded at 400 °C and failed catastrophically at 600 °C due to Rayleigh-Plateau instability.

**Conclusions**

In summary, we have demonstrated that the fundamental thermal instability of ultrathin metallic films can be overcome through van der Waals (vdW) epitaxy. Unlike conventional polycrystalline films that undergo stochastic grooving and rupture, vdW-templated films evolve toward flatter, more ordered morphologies that maintain structural continuity at temperatures close to the bulk melting point. This exceptional resilience originates from a restructured grain-boundary network dominated by low-energy, mechanically robust interfaces that suppress both interior rupture and edge retraction, even in the regime of rapid surface diffusion. By decoupling nanometric thickness from thermal fragility, vdW-textured metallic films provide a scalable architecture for resilient gates, interconnects, and contacts. This approach establishes a viable pathway for the high-temperature integration of 2D materials into next-generation high-frequency and high-power electronic systems.